\documentclass[twocolumn,showpacs,amsmath,amssymb,floatfix]{revtex4}

\usepackage{epsfig}
\usepackage{color}
\usepackage{times}
\usepackage{pst-all}
\usepackage{rotate}

\begin{document}

\title{Levels of Complexity in Scale-Invariant Neural Signals}

\author{Plamen~Ch.~Ivanov$^{1}$, Jeffrey M. Hausdorff$^2$, 
S. Havlin$^{3}$\\
Lu\'{\i}s~A.~Nunes~Amaral$^{1,4}$,  Kuniharu Arai$^5$, Verena~Schulte-Frohlinde$^1$\\
Mitsuru Yoneyama$^{6,7}$, H.~Eugene~Stanley$^1$}

\affiliation{$^{1}$ Center for Polymer Studies and Department of Physics, 
Boston University, Boston, MA 02215\\
$^{2}$Harvard Medical School, Boston, MA 02115 and Tel-Aviv Sourasky Medical Center, Tel-Aviv, Israel\\
$^3$ Department of Physics and Gonda-Goldschmied Center for Medical Diagnosis,
Bar-Ilan University, Ramat-Gan 52900, Israel\\
$^4$ Department of Chemical and Biological Engineering, Northwestern University, 
Evanson, Illinois 60208\\
$^5$ The Smith-Kettlewell Eye Research Institute, San Francisco, CA 94115\\
$^6$ Mitsubishi Chemical Group, Science and Technology Center Inc., Yokohama, Japan\\
$^7$ Research Institute for Electronic Science, Hokkaido University, Sapporo, Japan
}

\begin{abstract} 
  
Many physical and physiological signals exhibit complex scale-invariant
features characterized by $1/f$ scaling and long-range power-law
correlations, suggesting a possibly common control mechanism.
Specifically, it has been suggested that
dynamical processes influenced by inputs and feedback on multiple time scales
may be sufficient to give rise to $1/f$ scaling and scale invariance.  Two
examples of physiologic signals that are the output of hierarchical, multi-scale
physiologic systems under neural control are the human heartbeat and human
gait. Here we show that while both cardiac interbeat interval and gait
interstride interval time series under healthy conditions have comparable $1/f$
scaling, they still may belong to different complexity classes.  Our analysis
of the magnitude series correlations and multifractal scaling exponents of the
fluctuations in these two signals demonstrates that in contrast with the
nonlinear multifractal behavior found in healthy heartbeat dynamics, gait
time series exhibit less complex, close to monofractal behavior and a low
degree of nonlinearity. These findings 
underscore the limitations of traditional two-point correlation methods in
fully characterizing physiologic and physical dynamics. In addition, these results suggest
that different mechanisms of control may be responsible for varying levels of
complexity observed in physiological systems under neural regulation
and in physical systems that possess similar $1/f$ scaling.

\end{abstract}

\date{\today}  

\maketitle


\section{Introduction}

Many dynamic systems generate outputs with fluctuations characterized
by $1/f$-like scaling of the power spectra, $S(f)$, where $f$ is the
frequency. These fluctuations are often associated with nonequilibrium
dynamic systems possessing multiple degrees of freedom
\cite{Stanley71,bak94}, rather than being the output of a classic
``homeostatic'' process \cite{Bernard,van28,cannon29}.  It is generally assumed that
the presence of many components interacting over a wide range of time
or space scales could be the reason for the $1/f$ spectrum in the
fluctuations \cite{Johnson25,Dutta81}. Fluctuations exhibiting $1/f$-like
behavior are often termed ``complex'', since they obey a scaling law
indicating a hierarchical fractal organization of their frequency
(time scale) components rather than being dominated by a single
frequency.  $1/f$ behavior is common in a
variety of physical, biological and social systems
\cite{Dutta81,Vliet87,Weismann88,Musha77,Liu99,Hausdorff01,Ashkenazy02,Scafetta03,West03}.
The ubiquity of the $1/f$ scale-invariant phenomenon has triggered in recent
years the development of generic mechanisms describing complex systems,
independent of their particular context, in order to understand the
``unifying'' features of these systems
\cite{Shlesinger87,Shlesinger88,west89,Bassingthwaighte94}. 

To answer the question whether fluctuations in signals generated by integrated
physiological systems exhibit the same level of complexity, we
analyze and compare the time series generated by two physiologic
control systems under multiple-component integrated neural control --- the
human gait and the human heartbeat. We chose these two particular 
examples because human
gait and heartbeat control share certain fundamental properties, e.g., both
originate in oscillatory centers. In the case of the heart, the pacemaker is
located in the sinus node in the right atrium \cite{Berne96}. For gait,
pacemakers called central pattern generators are thought to be located in the
spinal cord \cite{Inman81}.

However, these two systems are distinct, suggesting possible dynamical
differences in their output. For example, heartbeat fluctuations are
primarily controlled by the involuntary (autonomic) nervous system. In
contrast, while the spontaneous walking rhythm is an automatic-like process,
voluntary inputs play a major role. Further, gait control resides in the
basal ganglia and related motor areas of the central nervous system, while
the heartbeat is controlled by the sympathetic and parasympathetic branches
of the autonomic nervous system \cite{Berne96,Levy71}.

Previous studies show comparable two-point linear correlations and $1/f$ power
spectra in heart rate \cite{Musha82,Yamamoto91,Yamamoto93,Peng93,Malik95} 
and human gait \cite{Kadaba89,Hausdorff95b,Yang02},
suggesting that differences in physiologic control may not be manifested in
beat-to-beat and interstride interval fluctuations. Recent studies focusing
on higher order correlations and nonlinear properties show that the human
heartbeat exhibits not only $1/f$ fractal but also multifractal properties
\cite{Ivanov99}. Since multifractal signals require many scaling indices to
fully characterize their scaling properties, they may be considered to be
more complex than those characterized by a single fractal dimension, such as
classical $1/f$ noise.  Although the origins of the multifractal features
in heartbeat dynamics are not yet understood, there is evidence that they
relate to the complex intrinsic neuroautonomic regulation of the heart
\cite{Ivanov99,Amaral01}. Human gait, e.g., free unconstrained walking, is
also a physiological process regulated by complex hierarchical feedback
mechanisms involving supra-spinal inputs
\cite{Inman81}. Moreover, recent findings indicate that the
scaling properties of gait fluctuations relate to neural centers on the
higher supra-spinal level rather than to lower motor neurons or environmental
inputs \cite{Collins93,Hausdorff96}. Thus it would be natural to hypothesize
that the fluctuations in healthy unconstrained human gait exhibit similar
fractal and multifractal features, and that human gait dynamics may belong to
the same ``complexity class'' as cardiac dynamics.

We employ two techniques --- magnitude and sign decomposition
analysis \cite{Ashkenazy-PRL01,Ashkenazy-PhysA-magn03} and multifractal
analysis \cite{Muzy91,Muzy94} --- to probe long-term nonlinear features, and to
compare the levels of complexity in heartbeat and interstride interval
fluctuations.
To this end, we analyze interstride interval time series from 10 young healthy
men (mean age 22 years) with no history of neuromascular disorders
\cite{Physiobank}. Subjects walked continuously for 1 hour at a self-selected
usual pace on level ground around a flat, obstacle-free, approximately oval,
400m long path. The interstride interval was measured using a ground reaction
force sensor 
--- ultra-thin force-sensitive switches were taped inside
one shoe and data were recorded on an ambulatory recorder using a previously
validated method \cite{Hausdorff95a}. We compare the results of our gait
analysis with results we have previously obtained
\cite{Ivanov99,IvanovEPL99,Ashkenazy-PRL01,Ivanov-Chaos01} from 6-hour long heartbeat
interval records from 18 healthy individuals (13 female and 5 male, mean age
34 years) during daily activity (12:00 to 18:00) \cite{Physiobank}.

As described below, we systematically compare the scaling
properties of the fluctuations in human gait with those in the human
heartbeat using power spectral analysis, detrended fluctuation analysis (DFA), 
magnitude and sign decomposition analysis, 
and wavelet-based multifractal analysis, and we quantify linear
and nonlinear features in the data over a range of time scales.

\section{Methods}

\subsection{Detrended fluctuation analysis (DFA)}

The DFA method was developed because conventional fluctuation
analyses, such as power spectral, R/S and Hurst analysis  
cannot be reliably used to study nonstationary data \cite{PengPRE93,Hu01,Chen02}.  One
advantage of the DFA method is that it allows the detection of long-range
power-law correlations in noisy signals with embedded polynomial trends that
can mask the true correlations in the fluctuations of a signal. The DFA
method has been successfully applied to a wide range of research fields in
physics~\cite{jan99,vandewalle99,Ivanovameteo1999_12,Montanari2000,malamudjstatlaninfer1999},
biology~\cite{SVDFA1,SMDFA1,taqqu95,PengPRE93,solomdnafractal1995}, and
physiology~\cite{iyengaramjphsiolreg,
makikallioheartamjcardiol1999,bundesleep2000,Laitio2000}.

The DFA method involves the following steps~\cite{Peng94}:

({\it i})  Given the original signal 
$s(i)$, where $i=1,..,N_{max}$ and $N_{max}$ is the length of the signal,
we first form the profile function
$y(k)\equiv\sum_{i=1}^{k}[s(i)-\langle s \rangle]$, where $\langle s
\rangle$ is the mean. One can consider the profile $y(k)$ as the position 
of a random walk in one dimension after $k$ steps.

({\it ii}) We divide the profile $y(k)$  into non-overlapping segments of equal
length $n$.

({\it iii}) In each segment of length $n$, we
fit $y(k)$, using a polynomial function of order $\ell$ which represents the
polynomial {\it trend\/} in that segment. The $y$
coordinate of the fit line in each segment is denoted by $y_n(k)$. 
Since we use a polynomial
fit of order $\ell$, we denote the algorithm as DFA-$\ell$.

({\it iv}) The profile function $y(k)$ is detrended by subtracting the local
trend $y_n(k)$ in each segment of length $n$.  In DFA-$\ell$, trends of order
$\ell-1$ in the original signal are eliminated. Thus, comparison of the
results for different orders of DFA-$\ell$ allows us to estimate the type of
polynomial trends in the time series $s(i)$.

({\it v}) For a given segment of length $n$, the root-mean-square (r.m.s.)
fluctuation for this integrated and detrended signal $s(i)$ is
calculated:

\begin{equation}
 F(n)\equiv\sqrt{{1\over {N_{max}}}\sum_{k=1}^{N_{max}}[y(k)-y_n(k)]^2}.
\label{F2}
\end{equation}

({\it vi}) Since we are interested in how $F(n)$ depends on the segment
length, the above computation is repeated for a broad range of scales $n$.

A power-law relation between the average root-mean-square fluctuation
function $F(n)$ and the segment length $n$ indicates the presence of scaling:
\begin{equation}
F(n)~\sim~n^{\alpha}. 
\label{F3}
\end{equation}
Thus, the DFA method can quantify the temporal
organization of the fluctuations in a given signal $s(i)$ by a single scaling
exponent $\alpha$ --- a self-similarity parameter which represents the
long-range power-law correlation properties of the signal. If $\alpha=0.5$,
there is no correlation and the signal is uncorrelated (white noise); if
$\alpha < 0.5$, the signal is anti-correlated; if $\alpha~>0.5$, the signal
is correlated. The larger the value of $\alpha$, the stronger the
correlations in the signal.

For stationary signals with scale-invariant temporal organization, $F(n)$ is
related to the Fourier power spectrum $S(f)$ and to the autocorrelation function
$C(n)$. For such signals, 

\begin{equation}
S(f) \sim f^{-\beta},~\mbox{where}~[~\beta~=~2\alpha-1~]
\label{F4}
\end{equation}
and $\alpha$ is the DFA scaling exponent (Eq.~\ref{F3}) \cite{PengPRE93,Peng94}. 
Thus signals with $1/f$ scaling in the power spectrum
(i.e. $\beta=1$) are characterized by DFA exponent $\alpha=1$. If
$0.5<\alpha<1$, the correlation exponent $\gamma$ describes the
decay of the autocorrelation function~\cite{PengPRE93}:
\begin{equation}
C(n)\equiv\langle s(i)s(i+n) \rangle \sim n^{-\gamma},
~\mbox{where}~[~\gamma=2-2\alpha~]. 
\label{F5}
\end{equation}

\subsection{Magnitude and sign decomposition method}

Fluctuations in the dynamical output of physical and physiological systems
can be characterized by their magnitude (absolute value) and their direction
(sign). These two quantities reflect the underlying interactions in a given
system --- the resulting ``force'' of these interactions at each moment
determines the magnitude and the direction of the fluctuations.  Recent
studies have shown that signals with identical long-range correlations can
differ in the time organization of the magnitude and sign of the
fluctuations~\cite{Ashkenazy-PRL01}. To assess the information contained in
these fluctuations, the magnitude and sign decomposition method was
introduced~\cite{Ashkenazy-PRL01,Ashkenazy-PhysA-magn03}.
This method involves the following steps:

 ({\it i}) Given the original signal $s(i)$ we generate the increment series,
$\Delta s(i) \equiv s(i+1)-s(i)$. 

({\it ii}) We decompose the increment series into a magnitude
series $|\Delta s(i)|$ and a sign series ${\rm sgn}(\Delta s(i))$.

({\it iii}) To avoid artificial trends we subtract from the magnitude and sign
series their average.

({\it iv}) We then integrate both magnitude and sign series, because of 
limitations in the accuracy of the
detrended fluctuation analysis method (DFA) for estimating the scaling
exponents of anticorrelated signals ($\alpha <0.5$).

({\it v}) We perform a scaling analysis using 2nd order detrended
fluctuation analysis (DFA-2) on the integrated magnitude and sign series.

({\it vi}) To obtain the scaling exponents for the magnitude and sign series we
measure the slope of $F(n)/n$ on a log-log plot, where $F(n)$ is the
root-mean-square fluctuation function obtained using DFA-2, and $n$ is the scale.

Fluctuations following an identical $1/f$ scaling law can exhibit different
types of correlations for the magnitude and the sign --- e.g., a signal with
anticorrelated fluctuations can exhibit positive correlations in the
magnitude. Positive correlations in the magnitude series indicate that an
increment with large magnitude is more likely to be followed by an increment
with large magnitude. Anticorrelations in the sign series indicate that a
positive increment in the original signal is more likely to be followed by a
negative increment. Further, positive power-law correlations in the magnitude
series indicate the presence of long-term {\it nonlinear} features in the original
signal, and relate to the width of multifractal
spectrum~\cite{Ashkenazy-PhysA-magn03}. In contrast the sign series relates
to the {\it linear} properties of the original
signal~\cite{Ashkenazy-PhysA-magn03}.  The magnitude and sign decomposition
method is suitable to probe nonlinear properties in short nonstationary
signals, such as 1-hour interstride interval time series.

\subsection{Wavelet-based multifractal analysis}

Previously, analyses of the fractal properties of physiologic fluctuations
revealed that the behavior of healthy, free-running physiologic systems may
often be characterized as
$1/f$-like~\cite{Kitney80,Kitney82,Musha82,Yamamoto91,Yamamoto93,Peng93, Liebovitch94,Bassingthwaighte94,
Kurths95,Hausdorff95a,Hausdorff95b,Malik95,Hausdorff96,Goldberger96,Chen97,West98,Lowen99,Havlin-S-1999a,Griffin00,Stanley-H-2000b,Protsmman01}.  Monofractal signals (such as classical
$1/f$ noise) are homogeneous, i.e., they have the same scaling properties
throughout the entire signal \cite{Bunde2,Takayasu,Dewey,Barabasi,Stoev02}.
Monofractal signals can therefore be indexed by a single exponent: the Hurst
exponent $H$ \cite{Hurst}.

On the other hand, multifractal signals are nonlinear and inhomogeneous with
local properties changing with time. Multifractal signals can be decomposed
into many subsets characterized by different {\it local\/} Hurst exponents
$h$, which quantify the local singular behavior and relate to the local
scaling of the time series. Thus, multifractal signals require many exponents
to fully characterize their properties~\cite{Stanley,Y16,Feder}. The
multifractal approach, a concept introduced in the context of multi-affine
functions \cite{VicsekTurbo91,Barabasi-A-1991a,Nittmann85,Meneveau87}, 
has the potential
to describe a wide class of signals more complex than those characterized by
a single fractal dimension.

The singular behavior of a signal $s(t)$ at time $t_{0}$ ---
${|s(t)-P_n(t)|\sim |t-t_0|^{h(t_0)}}$ for $t \rightarrow t_0$ --- is
characterized by the local Hurst exponent $h(t_0)$ where $n < h(t_0) < n+1$
and $P_n(t)$
is a polynomial fit of order $n$.  To avoid an {\it ad hoc} choice of the
range of time scales over which the local Hurst exponent $h$ is estimated,
and to filter out possible polynomial trends in the data which can mask local
singularities, we implement a wavelet-based algorithm~\cite{Muzy94}.
Wavelets are designed to probe time series over a broad range of scales and
have recently been successfully used in the analysis of physiological
signals~\cite{Li93,Meste94,Senhadji95a,Karrakchou95b,Thakor93,Morlet95,Morlet93,Reinhardt96,Karrakchou95}. In
particular, recent studies have shown that the wavelet decomposition reveals
a robust self-similar hierarchical organization in heartbeat
fluctuations, with bifurcations propagating from large to small
scales~\cite{Ivanov96,Ivanov98,Ivanov-Chaos01}. To quantify hierarchical
cascades in gait dynamics and to avoid inherent numerical instability in the
estimate of the local Hurst exponent, we employ a ``mean-field'' approach ---
a concept introduced in statistical physics~\cite{Stanley71} --- which allows
us to probe the collective behavior of local singularities throughout an
entire signal and over a broad range of time scales.

We study the multifractal properties of interstride interval time series
by applying the {\it wavelet transform modulus maxima\/} (WTMM)
method~\cite{Muzy91,Muzy93,Muzy94} that has been
proposed as a mean-field generalized multifractal formalism for
fractal signals.  We first obtain the wavelet coefficient at time $t_{0}$
from the continuous wavelet transform defined as:

\begin{equation}
W_a(t_0)~\equiv~a^{-1} \sum_{t=1}^{N} s(t) \psi ((t-t_0)/a) \,,
\label{e.coef}
\end{equation}
where $s(t)$ is the analyzed time series, $\psi$ is the analyzing wavelet function, 
$a$ is the wavelet scale (i.e., time scale of the analysis), and $N$ is the
number of data points in the time series. For $\psi$ we use the third
derivative of the Gaussian, thus filtering out up to 
second order polynomial trends in the data. We then choose
the modulus of the wavelet coefficients at each point $t$ in the time series
for a fixed wavelet scale $a$. 

Next, we estimate the partition function
\begin{equation}
Z_q(a) \equiv \sum_i |W_a(t)|^q \,,
\label{e.part-func}
\end{equation}
where the sum is only over the maxima values of $|W_a(t)|$, and the powers
$q$ take on real values. By not summing over the entire set of wavelet
transform coefficients along the time series at a given scale $a$ but only
over the wavelet transform modulus maxima, we focus on the fractal structure
of the temporal organization of the singularities in the signal
\cite{Muzy93}.

We repeat the procedure for different values of the wavelet scale
$a$ to estimate the scaling behavior 
\begin{equation}
{Z_q(a) \sim a^{\tau(q)}}.  
\label{e.z-scaling}
\end{equation}
In analogy
with what occurs in scale-free physical systems, in which phenomena
controlled by the same mechanism over multiple time scales are characterized
by scale-independent measures, we assume that the scale-independent measures,
$\tau(q)$, depend only on the underlying mechanism controlling the system.
Thus by studying the scaling behavior of \mbox{$Z(a,q)\sim a^{\tau(q)}$} we
may obtain information about the self-similar (fractal) properties of the
mechanism underlying gait control.

For certain values of the powers $q$, the exponents $\tau(q)$ have familiar
meanings.  In particular, $\tau(2)$ is related to the scaling exponent
of the Fourier power spectra, $S(f)\sim 1/f^{ \beta}$, as $\beta = 2 +
\tau(2)$ \cite{Muzy94}.  For positive $q$, $Z_q(a)$ reflects the scaling of the
large fluctuations and strong singularities in the signal, while for negative $q$,
$Z_q(a)$ reflects the scaling of the small fluctuations and weak
singularities \cite{Vicsek,Takayasu,Feder}.  Thus, the scaling exponents
$\tau(q)$ can reveal different aspects of the underlying dynamics.


In the framework of this wavelet-based multifractal formalism, $\tau(q)$
is the Legendre transform of the singularity spectrum $D(h)$
defined as the Hausdorff dimension of the set of points $t$ in the
signal $s(t)$ where the local Hurst exponent is $h$.  Homogeneous
monofractal signals --- i.e., signals with a single local Hurst exponent $h$
--- are characterized by linear $\tau(q)$ spectrum:  
\begin{equation}
\tau(q) = qH - 1, 
\label{e.tau-linear}
\end{equation}
where $H \equiv h={d\tau(q)}/{dq}$ is the global Hurst exponent. 
On the contrary, a nonlinear $\tau(q)$ curve is
the signature of nonhomogeneous signals that display 
multifractal properties---i.e., $h(t)$ is a varying quantity that
depends upon $t$.

\section{Results}

\begin{figure}[b!]
\centerline{
\epsfysize=0.95\columnwidth
{\rotate[r]{\epsfbox{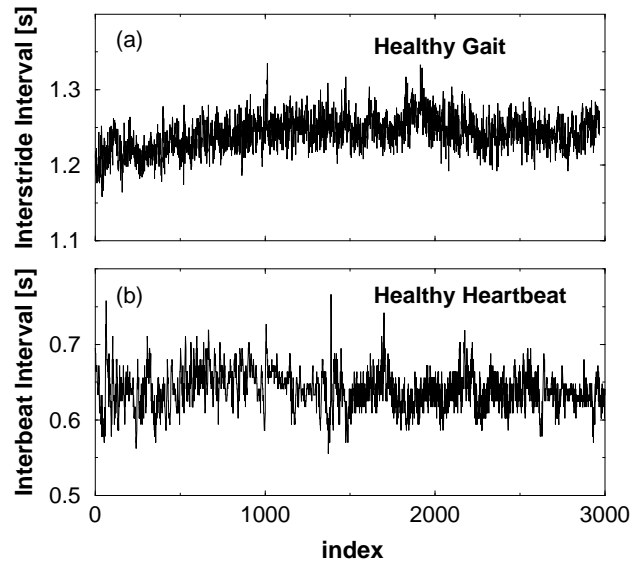}}}
}
\caption{
Representative records of (a) interstride interval time series from a healthy
subject and (b) consecutive heartbeat intervals from a healthy subject}
\label{data}
\end{figure}

In Fig.~\ref{data} we show two example time series: (i) an interstride
interval time series from a typical healthy subject during $\approx 1$ hour 
($3,000$ steps) of unconstrained normal walking on a level, obstacle-free
surface (Fig.~\ref{data}a) \cite{Physiobank}; (ii) consecutive heartbeat
intervals from $\approx 1$ hour ($3,000$ beats) record of a typical healthy
subject during daily activity (Fig.~\ref{data}b) \cite{Physiobank}. Both time
series exhibit irregular fluctuations and nonstationary behavior
characterized by different local trends; in fact it is difficult to
differentiate between the two time series by visual inspection.

We first examine the two-point correlations and scale-invariant behavior of
the time series shown in Fig.~\ref{data}. Power spectra $S(f)$ of the gait
and heartbeat time series (Fig.~\ref{correlations}a)
indicate that both processes are described by a power-law relation $S(f) \sim
1/f^{\beta}$ over more than 2 decades, with exponent $\beta \approx 1$. This
scaling behavior indicates self-similar (fractal) properties of the data
suggestive of an identical level of complexity as quantified by this linear
measure. We obtain similar results for the interstride interval times series
from all subjects in our gait database: $\beta=0.9 \pm 0.08$ 
(group mean $\pm$ std. dev.) in agreement with previous results \cite{Hausdorff96}.


\subsection{Detrended fluctuation analysis (DFA)}

\begin{figure}[th!]
\centerline{
\epsfysize=0.63\columnwidth{{\epsfbox{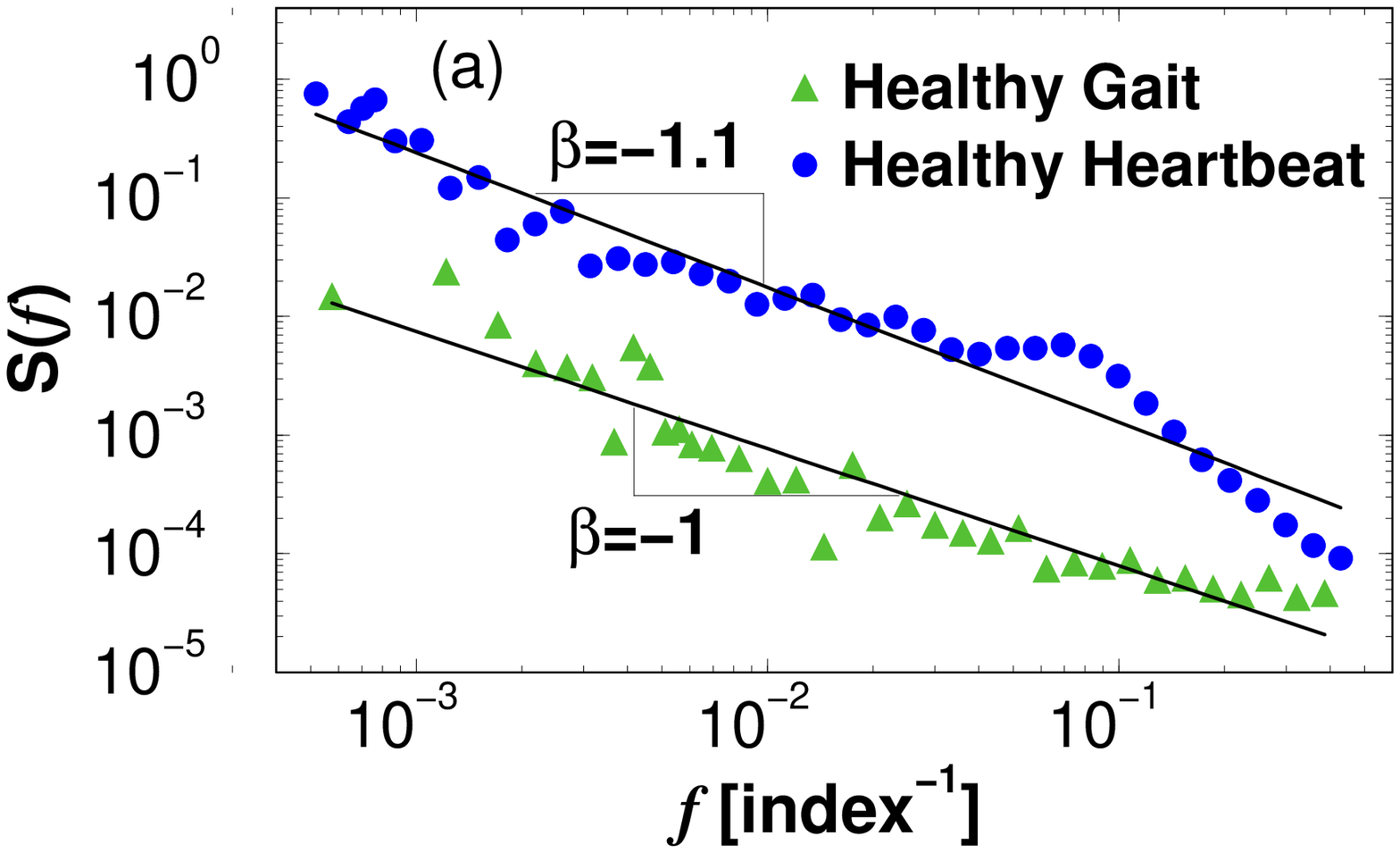}}}
}
\centerline{
\epsfysize=0.63\columnwidth{{\epsfbox{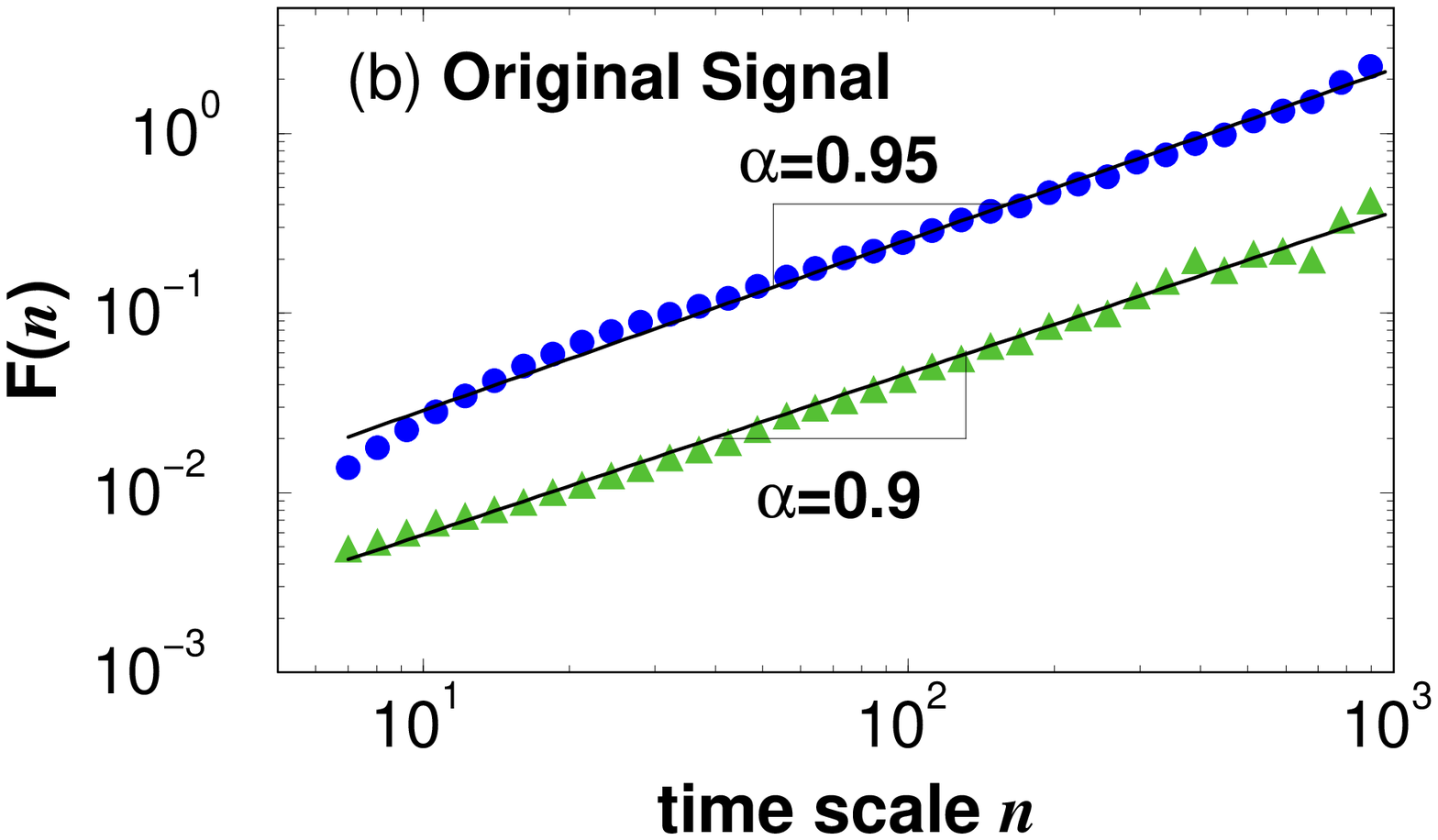}}}
}
\centerline{
\epsfysize=0.63\columnwidth{{\epsfbox{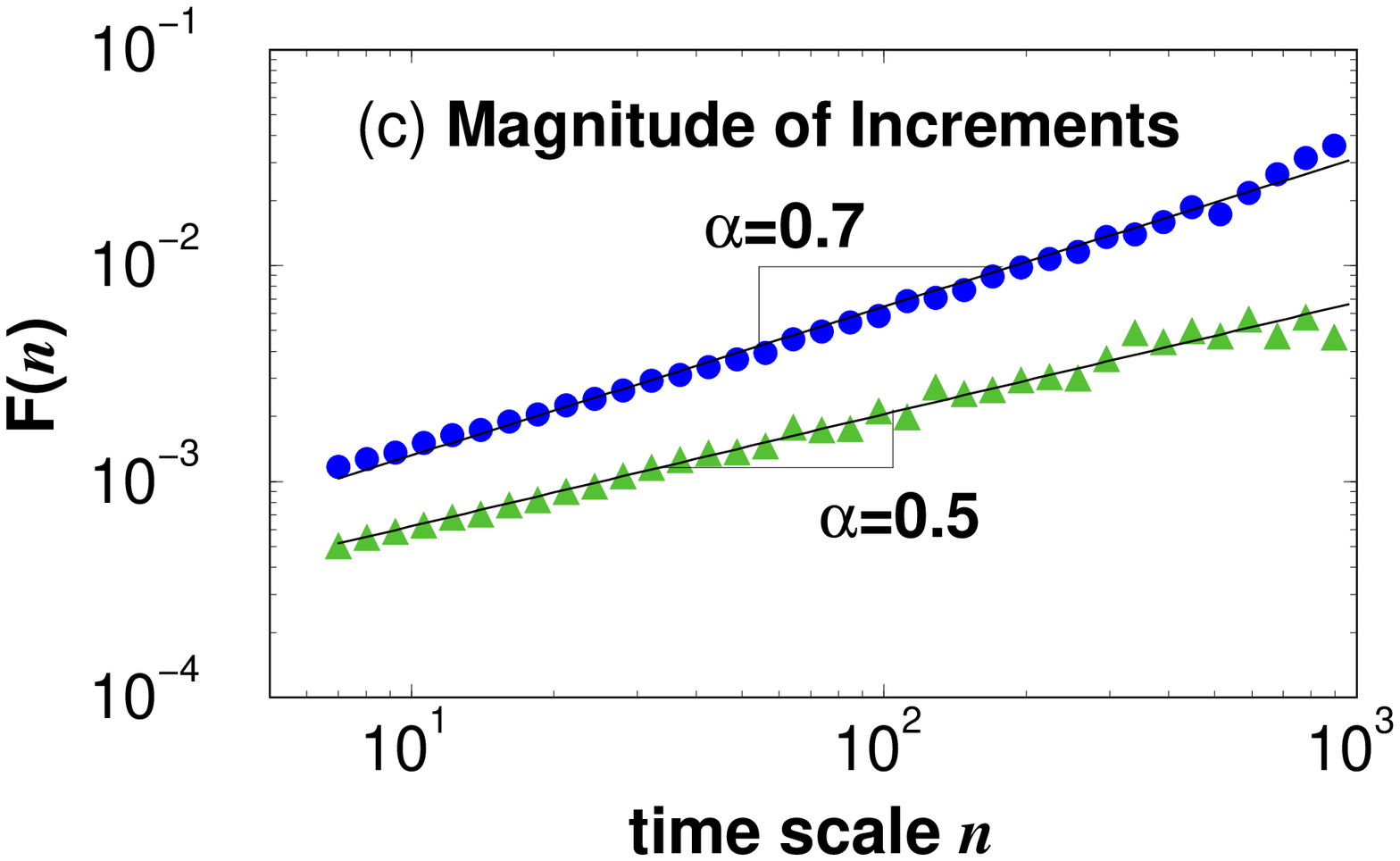}}}
}
\caption{
(a) Power spectra of the gait series ($\blacktriangle$) and 
heartbeat series ($\bullet$) displayed in
Fig.~\ref{data}. Plots of the root-mean-square fluctuation function $F(n)$
{\it vs.} time scale $n$ (measured in stride or beat number) from
second-order DFA-2 analysis for (b) the interstride and heartbeat interval time
series indicating similar power-law correlations, and (c) the magnitude
series of the interstride and heartbeat increments showing a surprising
difference in the nonlinear properties of the two time series.  }
\label{correlations}
\end{figure}

Next, to quantify the degree of correlation in the interstride and heartbeat
fluctuations we apply the DFA method, which also provides a linear measure:
plots of the root-mean-square fluctuation function $F(n)$ {\it vs.} time
scale $n$ (measured in stride or beat number) from a second-order DFA
analysis (DFA-2) \protect\cite{Peng94,Hu01,Chen02} indicate the presence of
long-range power-law correlations in both gait and heartbeat fluctuations
(Fig.~\ref{correlations}b).  The scaling exponent $\alpha \approx 0.95$ for
the heartbeat signal is very close to the exponent $\alpha \approx 0.9$ for
the interstride interval signal, estimated over the scaling range $6~<~n~<~600$. 
We obtain similar results for the remaining
subjects: $\alpha=0.87 \pm 0.03$ (group mean $\pm$ std. dev.) for the gait
data and $\alpha=1.04 \pm 0.08$ for the heartbeat data, in agreement with \cite{Hausdorff96}.

The results of both power spectral analysis and the DFA method indicate that
gait and heartbeat time series have similar scale-invariant properties
suggesting parallels in the underlying mechanisms of neural regulation.

\subsection{Magnitude and sign decomposition method}

To probe for long-term nonlinear features in the dynamics of interstride
intervals we employ the magnitude and sign decomposition analysis
\cite{Ashkenazy-PRL01,Ashkenazy-PhysA-magn03}. Previous studies have
demonstrated that information about the nonlinear properties of heartbeat
dynamics can be quantified by long-range power-law correlations in the
magnitude of the increments in heartbeat intervals
\cite{Ashkenazy-PRL01}. Further, 
correlations in the magnitude are
associated with nonlinear features in the underlying dynamics, while linear
signals are characterized by an absence of correlations (random behavior) in
the magnitude series. To quantify the correlations in the magnitude of the
interstride increments we apply the DFA-2 method to the data displayed in
Fig.~\ref{data}a. Our results show that the magnitude series of the
interstride increments exhibits close to random behavior with correlation
exponent $\alpha \approx 0.5$ 
(denoted by ($\blacktriangle$) in Fig.~\ref{correlations}c). 
In contrast, the
magnitude series of the heartbeat increments (Fig.~\ref{data}b) exhibits
strong positive correlations over more than two decades characterized by
exponent $\alpha=0.7$ (denoted by ($\bullet$) in Fig.~\ref{correlations}c). 
A surrogate test \cite{Panter65,Theiler92} eliminating the nonlinearity in the 
heartbeat time series by randomizing the
Fourier phases but preserving the power spectrum leads to random behavior
($\alpha=0.5$) in the magnitude series \cite{Ashkenazy-PRL01}. Thus the
striking difference in the magnitude correlations of gait and heartbeat
dynamics (both of which are under multilevel neural control) raises the
possibility that these two physiologic processes belong to different classes
of complexity whereby the neural regulation of the heartbeat is inherently
more nonlinear, over a range of time scales, than the neural mechanism of 
gait control. Our observation of
a low degree of nonlinearity in the gait time series is supported by the
remaining subjects in the group: over time scales $6~<~n~<~600$, 
we obtain exponent $\alpha_{\rm mag}=0.57 \pm 0.03$ 
(group mean $\pm$ std. dev.) for the gait time series, which is
significantly lower than the corresponding exponent $\alpha_{\rm mag}=0.75 \pm
0.06$ obtained for the heartbeat data ($p=4.8 \times 10^{-6}$, by the
Student's t-test).

\subsection{Wavelet-based multifractal analysis}

To further test the long-term nonlinear features in gait dynamics we 
study the multifractal properties of interstride time series. We apply
the Wavelet Transform Modulus Maxima (WTMM) method \cite{Muzy93,Muzy94} 
--- a ``mean-field'' type approach
to quantify the fractal organization of singularities in the signal. We
characterize the multifractal properties of a signal over a broad range of
time scales by the multifractal spectrum $\tau(q)$.

\begin{figure}[th!]
\centerline{
\epsfysize=0.66\columnwidth{{\epsfbox{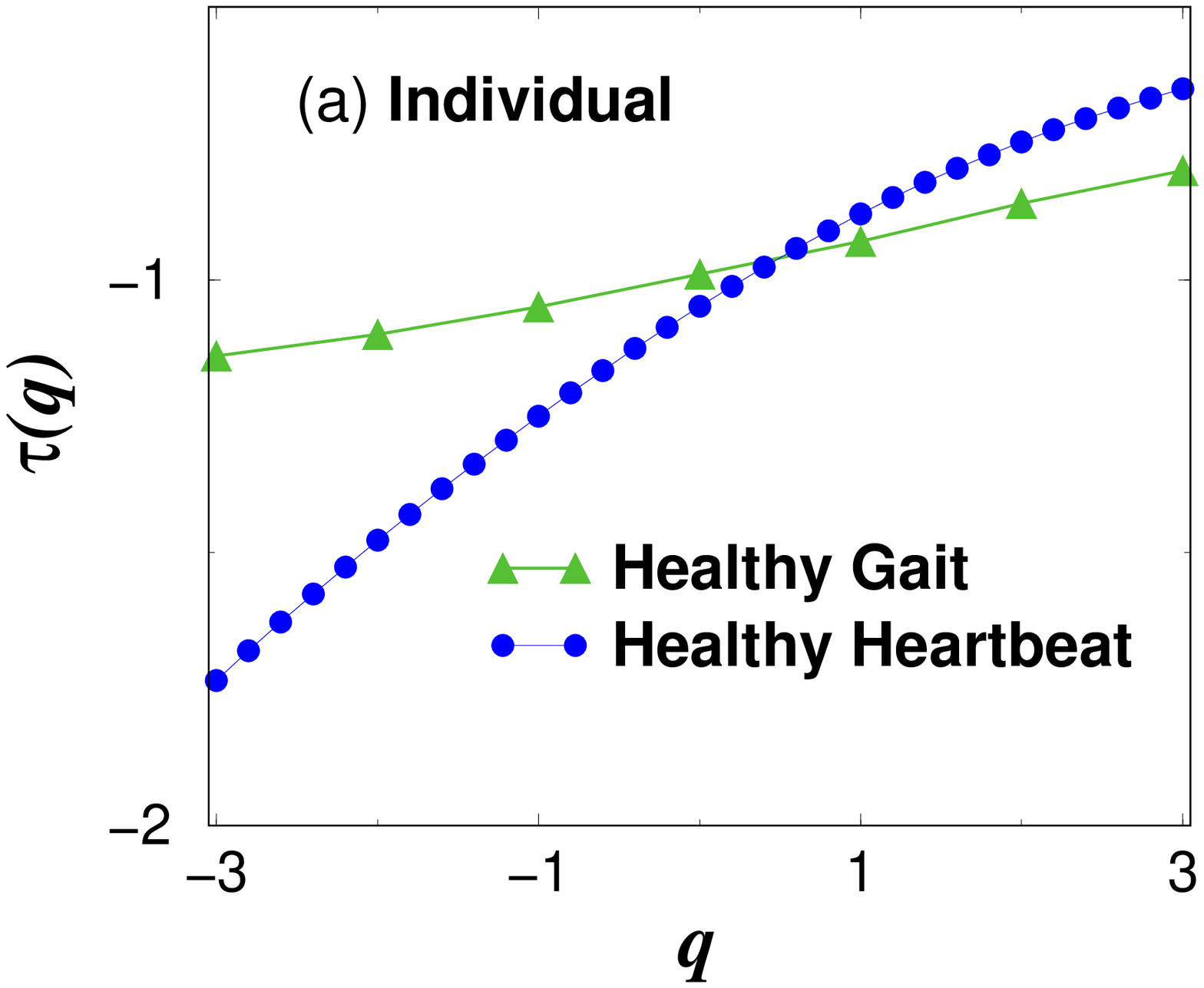}}}
}

\centerline{
\epsfysize=0.66\columnwidth{{\epsfbox{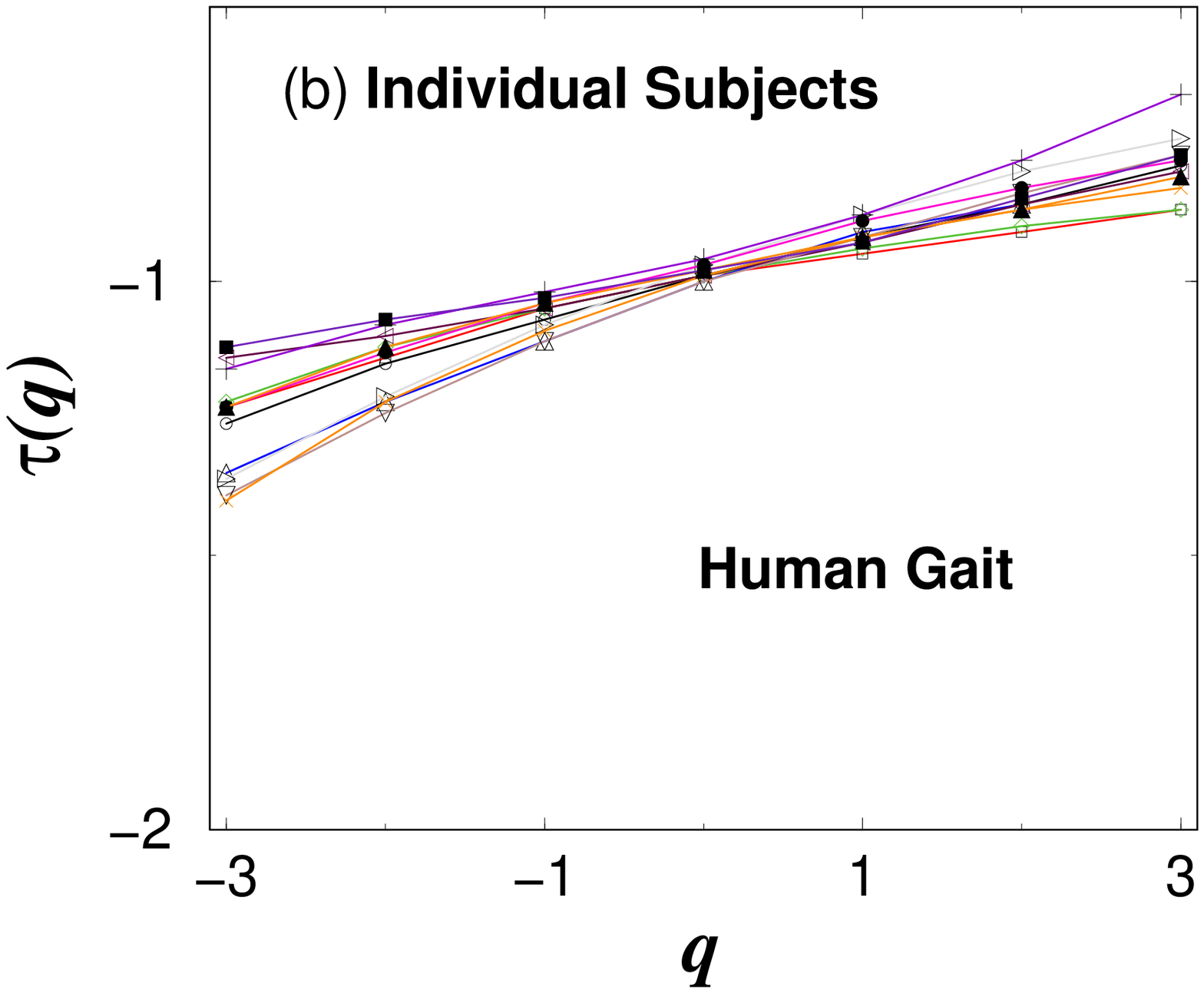}}}
}
\centerline{
\epsfysize=0.66\columnwidth{{\epsfbox{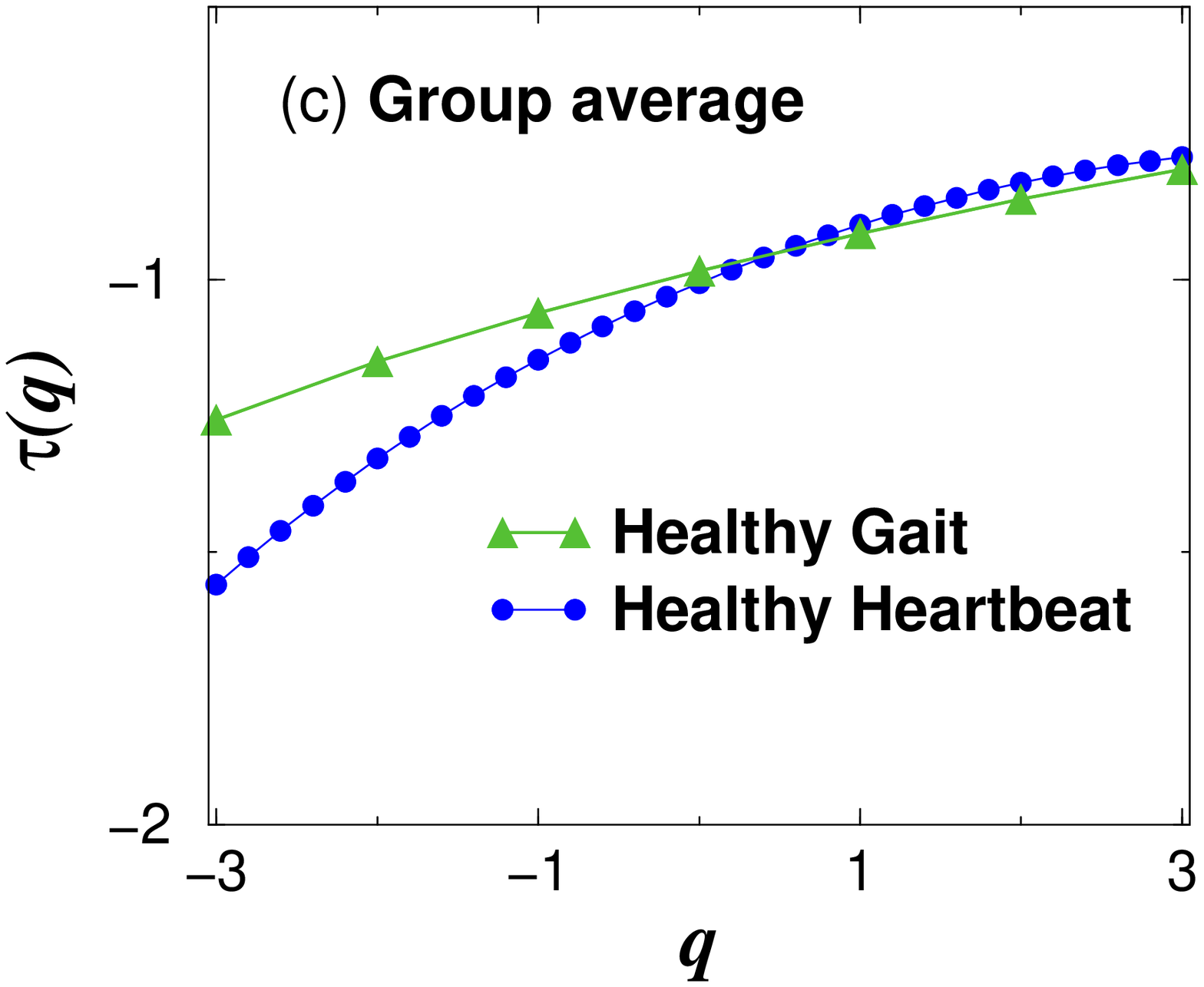}}}
}
\caption{ Multifractal analysis:
(a) Multifractal spectrum $\tau(q)$ for the individual
records shown in Fig.~\ref{data}, where $\tau$ is a scaling
index associated with different moments $q$ (Eq.~\ref{e.z-scaling}). A monofractal signal
corresponds to a straight line for $\tau(q)$, while for multifractal signals
$\tau(q)$ is a nonlinear function of $q$. The values of $\tau(q=2)$ for both 
gait and heartbeat time series are very close, in agreement with our findings 
based on DFA-2 correlation analysis (Fig.~\ref{correlations}b).
(b) Multifractral spectra $\tau(q)$ for all ten subjects in our database 
\cite{Physiobank} exhibit close to linear dependence on the moment $q$, 
suggesting monofractal behavior, in contrast to the nonlinear $\tau(q)$
spectra reported for heartbeat recordings \protect\cite{Stanley99}.
(c) Group average multifractal spectra $\tau(q)$ for the gait and heartbeat
subjects in our database \protect\cite{Physiobank}. The results
show a consistent monofractal (almost linear) behavior for the gait time
series, in contrast with the multifractal behavior
of the heartbeat data.
}
\label{multifractal}
\end{figure}

We first examine the time series shown in Fig.~\ref{data}. For the gait time
series, we obtain a $\tau(q)$ spectrum which is practically a linear function
of the moment $q$, suggesting that the gait dynamics exhibit {\it
monofractal} properties (Fig.~\ref{multifractal}a). This is in contrast with
the nonlinear $\tau(q)$ spectrum for the heartbeat signal
(Fig.~\ref{multifractal}a) which is indicative of multifractal behavior
\cite{Muzy91,Muzy94}. Further when analyzing the remaining interstride
interval recordings we find close to linear $\tau(q)$ spectra for all
subjects in the gait group (Fig.~\ref{multifractal}b). Calculating the group
averaged $\tau(q)$ spectra we find clear differences: multifractal behavior
for the heartbeat dynamics and practically monofractal behavior for the gait
dynamics (Fig.~\ref{multifractal}c).  Specifically we find significant
differences between the gait and heartbeat $\tau(q)$ spectra for negative
values of the moment $q$; for positive values of $q$, the scaling exponents
$\tau(q)$ take on similar values. This is in agreement with the similarity in
power spectral and DFA scaling exponents for gait and heartbeat data, which
correspond to $\tau(q=2)$ (Fig.~\ref{correlations}). 
However, the heartbeart $\tau(q)$ spectrum is visibly more curved for all 
moments $q$ compared with the gait $\tau(q)$ spectrum which may be approximately 
fit by a straight line, indicative of a low degree of nonlinearity in the 
interstride time series. Thus our results show consistent differences between the
nonlinear and multifractal properties of gait and heartbeat time series.

Previous studies have shown that reducing the level of physical activity
under a constant routine protocol does not change the multifractal features of
heartbeat dynamics, while blocking the sympathetic or parasympathetic tone of
the neuro-autonomic regulation of the heart dramatically changes the
multifractal spectrum, thus suggesting that the observed features in cardiac
dynamics arise from the intrinsic mechanisms of control \cite{Amaral01}.  Similarly,
by eliminating polynomial trends in the interstride interval time series
corresponding to changes in the gait pace using DFA and wavelet analyses, we
find scaling features which remain invariant among individuals. Therefore,
since different individuals experience different extrinsic factors, the
observed lower degree of nonlinearity as measured by the magnitude scaling
exponent and the close-to-monofractal behaviour characterized by practically
linear $\tau(q)$ spectrum appear to be a result of the intrinsic mechanisms
of gait regulation.  These observations suggest that while both gait and
heartbeat dynamics arise from layers of 
neural control
with multiple component interactions, and exhibit temporal organization over
multiple time scales, they nonetheless belong to different complexity classes. 
While both gait and heartbeat dynamics 
may be a result of competing inputs interacting through multiple
feedback loops, differences in the nature of these interactions may be
imprinted in their nonlinear and multifractal features: 
our findings suggest that while these interactions in heartbeat
dynamics are of a nonlinear character and are represented by Fourier phase
interactions encoded in the magnitude scaling and the multifractal spectrum, 
feedback mechanisms of gait
dynamics lead to decreased interactions among the Fourier phases.

\subsection{Further validation of gait results}

These new findings are supported by our analysis of a second group of gait
subjects. We analyze interstride intervals from an additional group of 7 young healthy
subjects (6 male, 1 female, mean age 28 years) recorded using a portable
accelerometer~\cite{japan-paper}.  Subjects walked continuously for $\approx 1$ hour at
a self-selected pace on an unconstrained outdoor walking track 
in a 
park environment allowing for slight changes
in elevation and obstacles related to pedestrian traffic. The stride interval
time series in this case were obtained from peak-to-peak intervals in the
accelerometer signal output in the direction of the subjects' vertical
axis~\cite{footnote}. Compatibility of the ground reaction force sensor used
for the gait recordings of the first group \cite{Hausdorff95a} with the
accelerometer device, and strong correlation between outputs of the two
devices was reported in Ref.~\cite{japan-paper}.

We find that for this second group the two-point correlation exponent
$\alpha$, as measured by the DFA method $\alpha=0.90 \pm 0.1$ (group mean
$\pm$ std. dev.) is similar to the group
average exponent of the first gait group ($\alpha=0.87 \pm 0.03$) and also
the heartbeat data ($\alpha=1.04 \pm 0.08$). In contrast, we find again a
significantly lower degree of nonlinearity, as measured by the magnitude
exponent $\alpha_{\rm mag}=0.62 \pm 0.04$ and the $\tau(q)$ spectrum, compared
with heartbeat dynamics $\alpha_{\rm mag}=0.75 \pm 0.06$ 
($p=1.3 \times 10^{-3}$, by the Student's t-test)
(Fig.~\ref{correlations}c and Fig.~\ref{multifractal}c).  On the other hand, the
group averaged value of $\alpha_{\rm mag}$ is slightly higher compared with the
first gait group ($\alpha_{\rm mag}=0.57 \pm 0.03$), and this is associated with
slightly stronger curvature in the $\tau(q)$ spectrum for the second gait group. This
may be attributed to the fact that the second group walked in a natural park
environment where obstacles, changes in elevation and pedestrian traffic may
possibly require the activation of higher neural centers.


The present results are related to a physiologically-based model of
gait control where specific interactions between neural centers are
considered \cite{Hausdorff01,Ashkenazy02}. 
In this model a lower degree of
nonlinearity (and close-to-linear monofractal $\tau(q)$ spectrum) 
reflects increased connectivity between neural centers, typically associated with
maturation of gait dynamics in adults. The present results are also consistent with
studies that used a different approach to quantify the dynamics of giat,
based on estimates of the
local Hurst exponents, and reported only weak multifractality in gait
dynamics \cite{Scafetta03,West03}.


\section{Summary}

In summary, we find that while the fluctuations in the output of both gait
and heartbeat processes are characterized by similar two-point correlation
properties and $1/f$-like spectra, they belong to different classes of
complexity --- human gait fluctuations exhibit linear and close to
monofractal properties characterized by a single scaling exponent, while
heartbeat fluctuations exhibit nonlinear multifractal properties which in
physical systems have been connected with turbulence and related multiscale
phenomena \cite{Nittmann85,Meneveau87,Muzy91,Frisch95}.

These findings are of interest because they underscore the limitations of
traditional two-point correlation methods in characterizing physiologic and physical time
series. In addition, these results suggest that feedback on multiple time
scales is not sufficient to explain different types of $1/f$ scaling and
scale-invariance, and highlight the need for the development of new models 
\cite{Ivanov-PhysA98,Lin01,McClintock02,Yulmetyev02}
that could account for the scale-invariant outputs of different types of
feedback systems.



\section{Acknowledgments}
We thank Y. Ashkenazy, A.L. Goldberger, Z. Chen, K. Hu and A. Yuen for helpful 
discussions and technical assistance.
This work was supported by grants from
NIH/National Center for Research Resources (P41 RR13622),
NSF, US-Israel BiNational Science Foundation and Mitsubishi Chemical Co., Yokahama, Japan. 



\newpage




\begin{thebibliography}{99}
\bibitem{footnote} 
The accelerometer device we used ($9\times6\times2cm$,
weight $140g$) was developed by Sharp Co. The device, attached to subjects' back, 
measures the vertical and anteroposterier acceleration profile during
walking. The output signals are digitized at a sampling frequency of,
$10^{3}Hz$, and are stored on a memory card. When the subjects' heel strikes
the ground a clear peak in the acceleration along the vertical axis is
recorded. The positions of these peaks in time are also controlled and
verified independently through matching steepest points in the
anteroposterier acceleration signal output.
\end{thebibliography}
\end{document}